\documentclass[aps,prb,superscriptaddress,reprint,floatfix,longbibliography,amsfonts,amsmath,amssymb]{revtex4-2}
\usepackage{graphicx}
\usepackage[colorlinks=true, urlcolor=blue, linkcolor=blue, citecolor=blue, pdftex]{hyperref}
\usepackage{epstopdf}
\usepackage{color}
\usepackage{pifont}
\usepackage{nicefrac}
\usepackage{array}
\usepackage[version=4]{mhchem}
\usepackage{bm}
\usepackage{pdfpages}
\makeatletter
\AtBeginDocument{\let\LS@rot\@undefined}
\makeatother
\newcommand{\fe}{\ce{Fe3Al2Si3O12}}
\newcommand{\mnge}{\ce{Mn3Al2Ge3O12}}
\newcommand{\gagaga}{\ce{Gd3Ga5O12}}
\newcommand{\TfnOneMultipole}{\(\mathbb{T}_{3}(A_{2u},1)\)}
\newcommand{\TfnOneMultipoleAbbrev}{\( \mathbb{T}(3,1) \equiv \mathbb{T}_{3}(A_{2u},1)\)}

\newcommand{\nh}{\hat{\bm n}}

\newcommand{\suppSecDFT}{Supplementary Note 1}
\newcommand{\suppSecStructure}{Supplementary Note 2}
\newcommand{\suppSecDSF}{Supplementary Note 3}
\newcommand{\suppSecNeutronFM}{Supplementary Note 4}
\newcommand{\suppSecMinimalModel}{Supplementary Note 5}
\newcommand{\suppSecDirector}{Supplementary Note 6}

\begin{document}
\title{Hyperkagome antiferromagnetism in the iron garnet Almandine \ce{Fe3Al2Si3O12}}
 
\author{Minnong Zhang}
\affiliation{Department of Physics, Okayama University, Okayama 700-8530, Japan}

\author{Harald O. Jeschke}
\email{jeschke@okayama-u.ac.jp}
\affiliation{Research Institute for Interdisciplinary Science, Okayama University, Okayama 700-8530, Japan}

\author{Daniel Guterding}
\email{guterding@th-brandenburg.de}
\affiliation{Technische Hochschule Brandenburg, Magdeburger Straße 50, 14770 Brandenburg an der Havel, Germany}

\date{\today}

\begin{abstract}
Antiferromagnets on the hyperkagome lattice provide a three-dimensional route to strong magnetic frustration, but experimentally accessible realizations with well-established microscopic Hamiltonians remain scarce. Garnets are a broad family of materials in which magnetic ions can form two interpenetrating hyperkagome lattices, giving rise to spin glass, spin-liquid and hidden-order physics. Here we identify the iron aluminosilicate garnet almandine, \ce{Fe3Al2Si3O12}, as a spin-$2$ hyperkagome antiferromagnet. Using density functional theory energy mapping, we determine a Heisenberg Hamiltonian for the high-spin Fe$^{2+}$ moments and find dominant nearest-neighbor antiferromagnetic exchange, as well as significant longer-range interactions. Classical Monte Carlo simulations of this Hamiltonian, combined with quantum-to-classical rescaling of the spin length, show a transition to antiferromagnetic order at $T_{\rm N}\approx9.8$\,K, in excellent agreement with specific heat experiments. The resulting non-collinear magnetic state is naturally described by magnetic toroidal octupolar order, which we connect to known results for the highly frustrated hyperkagome antiferromagnet gadolinium gallium garnet, {\gagaga}. We also calculate magnetic neutron diffraction amplitudes as well as static and dynamic structure factors using semiclassical dynamics, providing quantitative predictions for future neutron scattering experiments. Our results establish almandine as a chemically accessible realization of frustrated hyperkagome magnetism in the garnet family.
\end{abstract}

\maketitle

\section{Introduction}
\begin{figure}[t]
\centering
\includegraphics[width=0.95\linewidth]{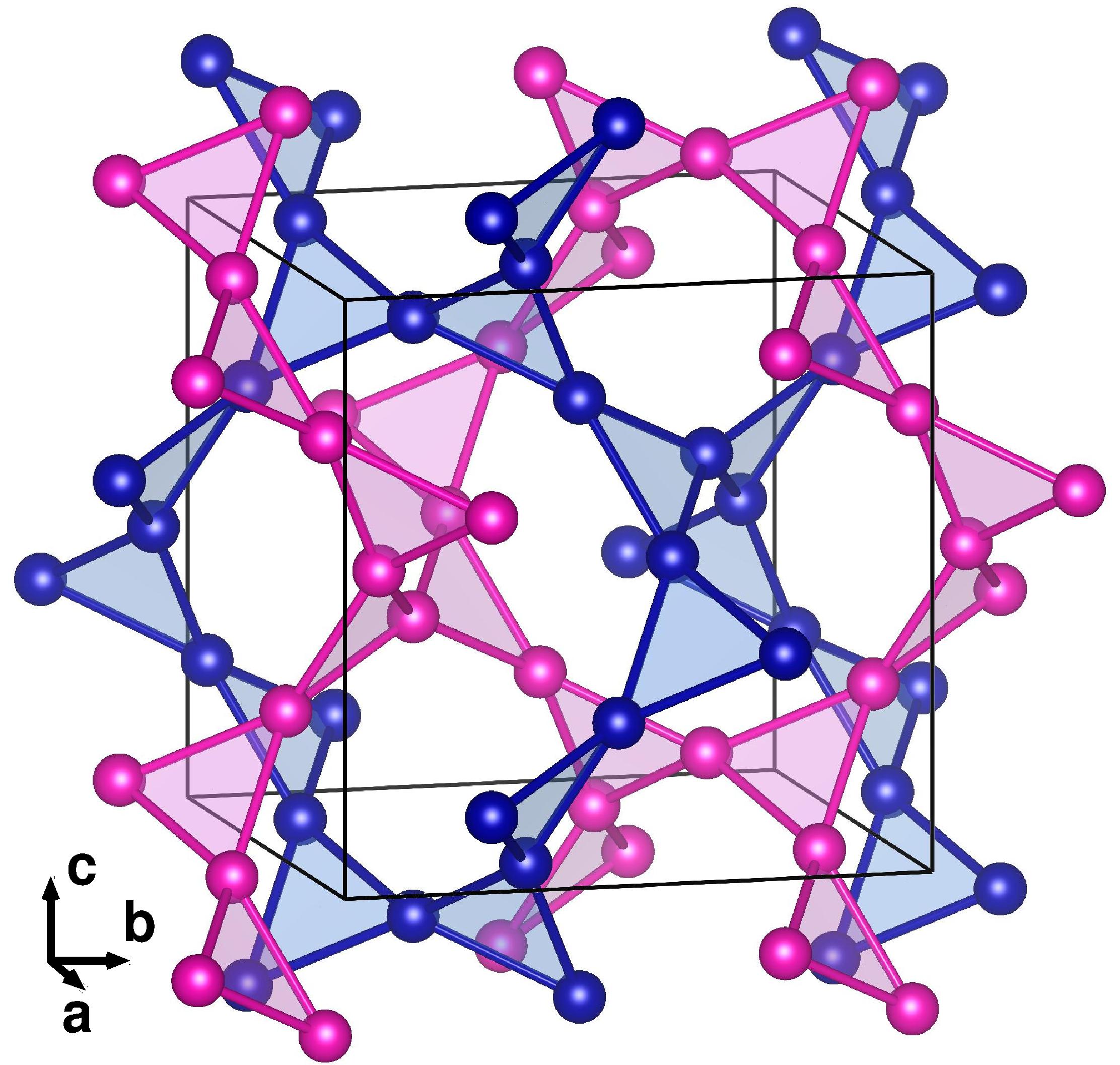}
\caption{\textbf{Representative garnet structure with two interpenetrating hyperkagome lattices.} The two hyperkagome lattices are respectively colored in blue and purple.}
\label{fig:garnet}
\end{figure}

The kagome lattice with corner-sharing triangles is at the heart of the study of frustrated magnetism. Removing one quarter of the sites in the pyrochlore lattice leads to a three-dimensional analogue of the kagome lattice, the hyperkagome lattice~\cite{Hopkinson2007}. The garnet structure has two interpenetrating hyperkagome lattices as shown in Fig.~\ref{fig:garnet}. In principle, magnetic hyperkagome lattices are widespread in nature: they form whenever magnetic $3d$ transition-metal or $4f$ rare-earth ions occupy the $24c$ site of the garnet space group $Ia\bar{3}d$. Arguably the most famous example in the field of strongly frustrated magnetism is gadolinium gallium garnet (GGG) {\gagaga}~\cite{Sasvari1983}. It has been analyzed as a spin glass~\cite{Schiffer1995} before it was described as a spin liquid~\cite{Petrenko1998} in which a hidden order was detected~\cite{Paddison2015}. 

The discovery of \ce{Na4Ir3O8}, where Ir$^{4+}$ ions realize $J_{\rm eff}=\frac{1}{2}$ magnetic moments on the so-called half-garnet lattice~\cite{Okamoto2007}, has led to intense theoretical efforts to understand the quantum hyperkagome antiferromagnet~\cite{Lawler2008,Lawler2008b,Zhou2008,Bergholtz2010}. Different kinds of quantum spin liquids (QSLs)~\cite{Lawler2008,Lawler2008b,Zhou2008} and their properties~\cite{Wan2016} as well as possible orders~\cite{Bergholtz2010} have been discussed.
A highly frustrated realization of the hyperkagome lattice was discovered in \ce{PbCuTe2O6}. Due to additional triangles added onto the regular hyperkagome lattice, it was termed hyper-hyperkagome~\cite{Chillal2020}. This material has attracted considerable theoretical attention~\cite{Chern2021}, since it is one of few good candidates for QSLs in three dimensions (besides \ce{Na4Ir3O8}).

\begin{figure*}
\centering
\includegraphics[width=0.95\textwidth]{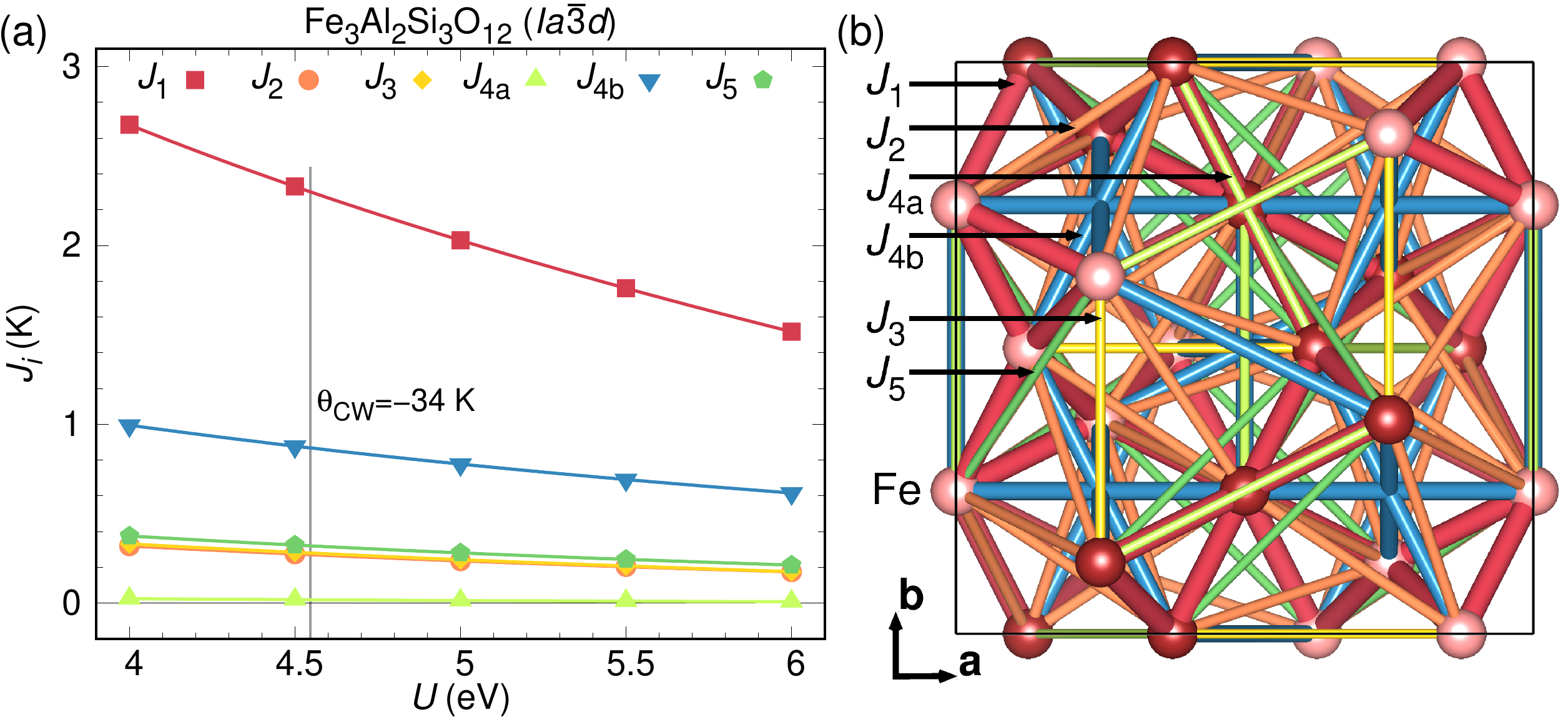}
\caption{\textbf{Heisenberg Hamiltonian of almandine {\fe}.} \textbf{a} Heisenberg couplings from DFT energy mapping for almandine {\fe}. Positive values correspond to antiferromagnetic exchange. The grey vertical line marks the on-site interaction $U$ which reproduces the experimental Curie-Weiss temperature $\theta_{\rm CW}$. \textbf{b} Exchange paths in the garnet structure up to the fifth neighbor. The two hyperkagome lattices are distinguished by lighter and darker Fe sites.}
\label{fig:couplings}
\end{figure*}

Although many garnets, including numerous natural minerals, carry magnetic ions on the $24c$ site, clean realizations of \textit{magnetic} hyperkagome lattices are rare: frequently the octahedral or tetrahedral sites are magnetic as well, or natural samples are solid solutions with substantial site mixing. Here, we would like to showcase the iron garnet almandine (\fe), which is a silicate mineral continuously studied for its important role in processes in the Earth's crust and mantle~\cite{Pownall2023}. Its structure has been known for almost a century~\cite{Menzer1929}, and its magnetism has been investigated with thermodynamic probes~\cite{deOliveira1989,Anovitz1993,Dachs2012}, with M\"ossbauer spectroscopy~\cite{Prandl1971,Murad1987} and neutron diffraction~\cite{Prandl1971}. Structural investigations with x-ray diffraction~\cite{Armbruster1992,Chaplot2002} as well as inelastic neutron scattering~\cite{Mittal2000} exist. Previous efforts to work out the Hamiltonian of almandine~\cite{Zherebetskyy2012} have not been very successful in explaining the experimental data.

In this work, we use density functional theory based energy mapping to show that almandine is a very interesting entry in the list of hyperkagome antiferromagnets; like in the paradigmatic gadolinium gallium garnet~\cite{Yavorskii2006}, the Hamiltonian is incomplete without subdominant but clearly nonzero longer range couplings. We use classical Monte Carlo simulations to show that in the light of existing experiments, our Hamiltonian is a very good candidate. The resulting magnetic ground state is non-collinear and naturally described by a magnetic toroidal octupolar order parameter. We connect this result to various order parameters that have been discussed in relation to gadolinium gallium garnet~\cite{Paddison2015}. In almandine these properties are, however, coupled to a conventional ordered spin-state. We discuss which aspects of the Hamiltonian lead to conventional non-collinear order in almandine.
In addition, we employ semiclassical dynamics~\cite{Remund2022} and linear spin-wave theory (LSWT) to study the static and dynamical spin structure factors of almandine and give predictions for the outcome of future neutron scattering experiments.

\section{Results}
\begin{table}[tb]
\caption{\textbf{Calculated Heisenberg exchange couplings for almandine {\fe}.} All values are given in units of Kelvin.
Parentheses indicate uncertainties in the last digit.}
\label{tab:couplings}
\begin{ruledtabular}
\begin{tabular}{cccccc}
$J_1$ & $J_2$ & $J_3$ & $J_{4a}$ & $J_{4b}$ & $J_5$ \\
\hline
2.30(1) & 0.27(1) & 0.28(1) & 0.02(1) & 0.87(1) & 0.32(1) \\
\end{tabular}
\end{ruledtabular}
\end{table}
We use density functional theory based energy mapping (for details see Section~\ref{sec:methodsdft}) to work out the Heisenberg Hamiltonian of {\fe}
\begin{equation}
H=\sum_{i<j} J_{ij}{\bf S}_i\cdot {\bf S}_j \,,
\label{eq:H}
\end{equation}
where ${\bf S}_i$ are spin operators of length $S=2$, and we do not double count bonds.
Our approach allows us to determine seventeen inequivalent exchange interactions up to about three times the nearest neighbor distance.
Fig.~\ref{fig:couplings}a shows the leading exchange couplings for five different values of the on-site interaction strength $U$ and fixed Hund's rule coupling $J_{\rm H}=0.80$\,eV~\cite{Mizokawa1996}. We determine the relevant Hamiltonian parameters by demanding that the set of couplings should yield the experimental Curie-Weiss temperature $\theta_{\rm CW}=-34$\,K~\cite{deOliveira1989}. The corresponding value of $U=4.55$\,eV is marked by a vertical line in Fig.~\ref{fig:couplings}a, and the result is $J_1= 2.30(1)$\,K, $J_2= 0.27(1)$\,K, $J_3= 0.28(1)$\,K, $J_{4a}= 0.02(1)$\,K, $J_{4b}= 0.87(1)$\,K, $J_5= 0.32(1)$\,K. These are summarized in Table~\ref{tab:couplings} (see {\suppSecDFT} for a table containing all calculated couplings). We find all exchange interactions to be antiferromagnetic. Two important aspects are the separation of $J_{4a}$ and $J_{4b}$ (see Fig.~\ref{fig:couplings}b), which have the same Fe-Fe distance but are inequivalent by symmetry, because only $J_{4b}$ has an Al$^{3+}$ ion in the middle of the path; $J_{4b}$ turns out to be the dominant sub-leading coupling. The second most important subdominant coupling is $J_5$, which is the third neighbor within each of the two hyperkagome lattices ($J_{4a}$ being the second), while $J_2$, $J_3$ and $J_{4b}$ are all inter-hyperkagome lattice couplings.

\begin{figure*}
\centering
\includegraphics[width=\linewidth]{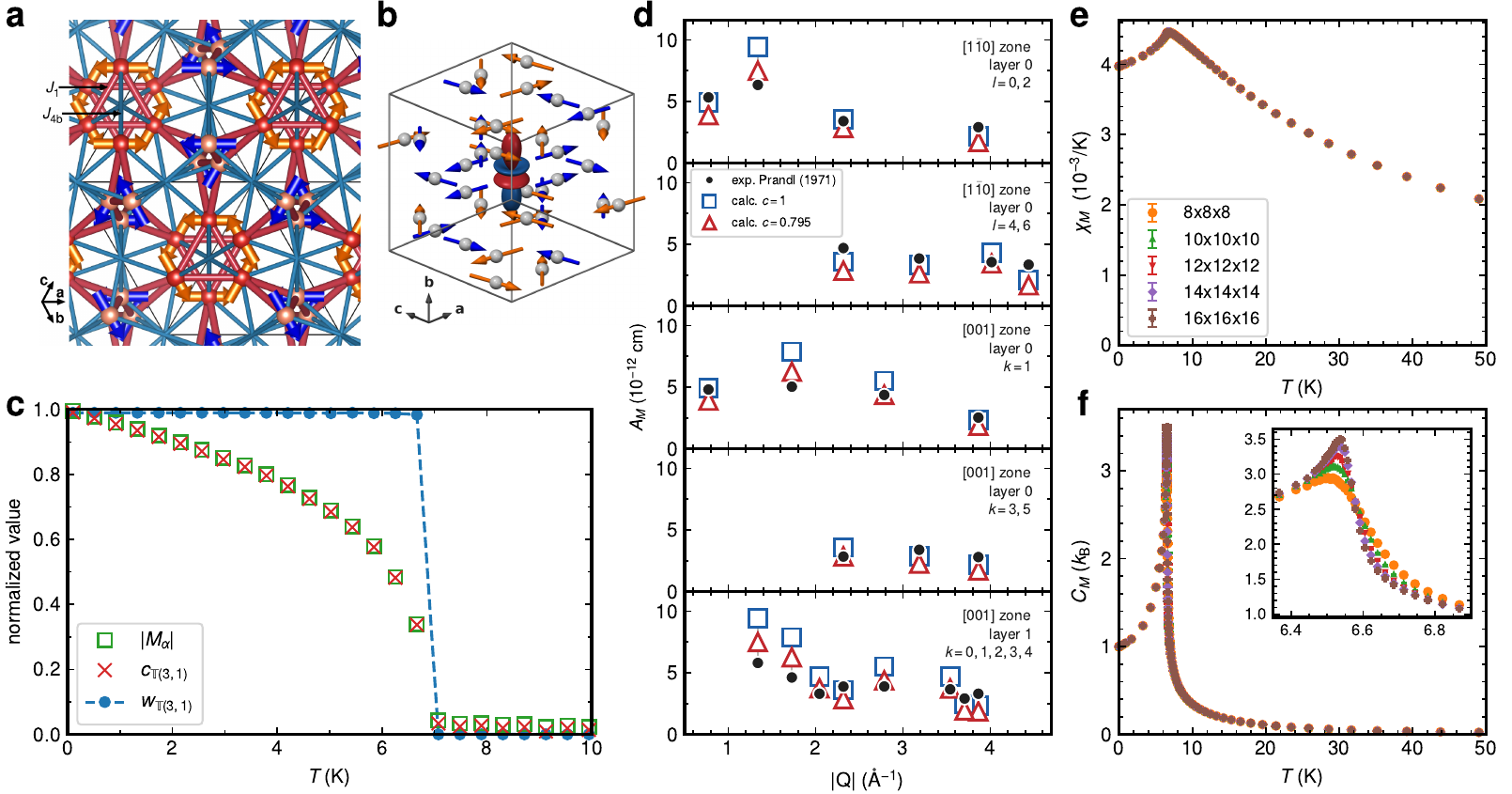}
\caption{
\textbf{Magnetic ground state and low-temperature thermodynamic properties of almandine.}
\textbf{a} Ground state magnetic order of {\fe} at $T=0.001$\,K. The magnetic space group is $R\bar{3}'c$ (BNS no.~167.105). In this rhombohedral space group, each of the two symmetry-inequivalent Fe sites occurs on both interpenetrating hyperkagome lattices, unlike in the parent space group $Ia\bar{3}d$.
\textbf{b} Ground state spin order and associated toroidal octupolar polynomial~\cite{Yatsushiro2021, Kusonose2023} for the {\TfnOneMultipoleAbbrev} multipole in the cubic unit cell.
\textbf{c} Dipolar order parameter (sublattice magnetization $|M_\alpha|$) and {\TfnOneMultipoleAbbrev} toroidal octupolar order parameter ($c_{{\scriptscriptstyle\mathbb{T}}(3,1)}$) as a function of temperature. $w_{{\scriptscriptstyle\mathbb{T}}(3,1)}$ denotes the relative weight of the toroidal octupolar order in the multipole expansion. At the transition temperature the overlap of the cMC spin configuration with the multipole expansion vanishes.
\textbf{d} Comparison of experimental~\cite{Prandl1971} and theoretical neutron diffraction results. Our calculations are based on the $R\bar{3}'c$ magnetic structure. $c=1$ denotes stoichiometric almandine, while $c=0.795$ simulates the fractional occupation of Fe positions assumed in Ref.~\cite{Prandl1971}.
\textbf{e} Magnetic susceptibility and \textbf{f} magnetic specific heat of almandine calculated from classical Monte Carlo.
}
\label{fig:groundstate}
\end{figure*}

We now investigate the properties of the above Hamiltonian for almandine using classical Monte Carlo (cMC) simulations (for details see Section \ref{sec:methodscmc}). We obtained the magnetic ground state from cMC at $T=0.001$\,K for the supercell with $N_{\rm s}=12288$ spins. Its magnetic space group is $R\bar{3}'c$ (BNS no.~167.105). Because Eq.~\ref{eq:H} is invariant under global spin rotations, it fixes this texture only up to a common rotation of all spins. A rotation by $90^\circ$ about the trigonal axis yields an equally symmetric structure with magnetic space group $R\bar{3}'c'$ (BNS no.~167.106). The dipole-dipole interaction lifts this degeneracy in favor of $R\bar{3}'c$, and our cMC simulations including dipole-dipole interactions order directly into $R\bar{3}'c$ (see {\suppSecStructure}). A sufficiently strong single-ion anisotropy of Fe$^{2+}$, which our model does not contain, could instead stabilize $R\bar{3}'c'$. The resulting real-space configuration in Fig.~\ref{fig:groundstate}a shows that the spins are concentrated around six well-defined directions. These directions form three antiparallel pairs, as well as two sets of spins at 120 degree angles. For the pure Heisenberg Hamiltonian in Eq.~\ref{eq:H}, all spins are coplanar. This can be easily seen to satisfy the two most important exchange interactions: AFM $J_1$ ensures a 120 degree state on every triangle of the hyperkagome lattice, and the AFM inter-hyperkagome coupling $J_{4b}$ forces the spins it connects into an antiparallel arrangement. The dipole-dipole interaction cants the spins slightly out of the plane perpendicular to the trigonal axis, by $1.2^\circ$ on Fe1 and by $0.3^\circ$ in the opposite sense on Fe2 (Table~\ref{tab:magneticstructure}). This staggered canting is allowed by $R\bar{3}'c$ and produces no net moment. Spins connected by $J_{4b}$ remain exactly antiparallel, and the angles on the $J_1$ triangles deviate from $120^\circ$ by less than $0.05^\circ$. Note that the two Fe sublattices, denoted as lighter and darker Fe atoms, occur on both hyperkagome lattices in space group $R\bar{3}'c$, unlike in the parent space group $Ia\bar{3}d$.

In Fig.~\ref{fig:groundstate}b we show the spin pattern in the parent cubic cell. We have shifted the origin so that the Al site, which usually sits in the corner, appears in the center of the cell. This shift is for display only. The multipole projection uses the convention given in Section~\ref{sec:methodsmultipole}. The magnetic ground state we determined earlier is perfectly described by a magnetic toroidal octupolar order~\cite{Yatsushiro2021, Kusonose2023} of {\TfnOneMultipole} type (see Section~\ref{sec:methodsmultipole}), centered on the Al site. The associated three-dimensional scalar polynomial is drawn in the center of the cell. Subsequently, we analyzed the magnetic state as a function of temperature. Thermal fluctuations introduce disorder, hence we expect that finite-temperature states deviate from the perfect toroidal octupolar order. Fig.~\ref{fig:groundstate}c shows the evolution of the dipolar (spin sublattice magnetization $|M_\alpha|$) and the projection ($c_{{\scriptscriptstyle\mathbb{T}}(3,1)}$) of actual finite-temperature spin states onto the {\TfnOneMultipoleAbbrev} toroidal octupolar state. $w_{{\scriptscriptstyle\mathbb{T}}(3,1)}$ tracks the weight of the {\TfnOneMultipole} state within the full multipole expansion. Fig.~\ref{fig:groundstate}c clearly shows that dipole and octupolar order develop concomitantly starting at the N{\'e}el temperature of about 6.5\,K. Within the ordered state, the multipole expansion is almost purely of {\TfnOneMultipole} type, independent of the specific temperature. For further details, see Section~\ref{sec:methodsmultipole}.

\begin{table}[t]
\caption{\label{tab:magneticstructure} \textbf{High-symmetry magnetic structure of {\fe}.} Atomic positions and ordered moments are given in the standard trigonal setting of the BNS 167.105 ($R\bar{3}'c$) magnetic structure obtained from our cMC calculations, with $a=b=16.264022\,\mathrm{\AA}$, $c=9.959639\,\mathrm{\AA}$, $\alpha = \beta = 90^\circ$ and $\gamma = 120^\circ$. Multiplicities refer to the full magnetic unit cell. Moment components are coefficients along the unit vectors of the non-orthogonal crystal axes, with $|\mathbf m|=4\,\mu_{\rm B}$. The origin is at the Al1 site, and the magnetic propagation vector is $k=(0,0,0)$.}
\begin{ruledtabular}
\begin{tabular}{lcrrr}
\multicolumn{5}{l}{Atomic positions} \\
Site & Mult. & $x$ & $y$ & $z$ \\
\hline
Fe1 & 18 & $\nicefrac{13}{24}$ & $\nicefrac{5}{24}$ & $\nicefrac{7}{12}$ \\
Fe2 & 18 & $\nicefrac{7}{24}$ & $\nicefrac{23}{24}$ & $\nicefrac{1}{12}$ \\
Al1 & 6 & 0 & 0 & 0 \\
Al2 & 18 & $\nicefrac{1}{3}$ & $\nicefrac{1}{6}$ & $\nicefrac{2}{3}$ \\
Si1 & 36 & $\nicefrac{7}{24}$ & $\nicefrac{11}{24}$ & $\nicefrac{1}{12}$ \\
O1 & 36 & 0.5754 & 0.3424 & 0.7221 \\
O2 & 36 & 0.4873 & 0.0949 & 0.4034 \\
O3 & 36 & 0.3935 & 0.0955 & 0.2158 \\
O4 & 36 & 0.2104 & 0.8006 & 0.9921 \\
\hline
\hline
\multicolumn{5}{l}{Ordered moments ($\mu_B$)} \\
Site & & $m_x$ & $m_y$ & $m_z$ \\
\hline
Fe1 & & 2.3089 & -2.3089 & -0.0856 \\
Fe2 & & 2.3094 & -2.3094 & 0.0217 \\
\end{tabular}
\end{ruledtabular}
\end{table}

Interestingly, we can connect this octupolar {\TfnOneMultipole} order to previously suggested order parameters for the pure hyperkagome lattice~\cite{Zhitomirsky2008} and gadolinium gallium garnet~\cite{Paddison2015}. Zhitomirsky's third-rank spin tensor detects the common orientation of three spin directions separated by $120^\circ$, without specifying their assignment to lattice sites. In almandine, the six spin directions form two approximately opposite triads, whose uniform spin-octupolar tensors cancel in the ideal limit. We therefore use a staggered combination to characterize their triatic orientational order (see Section~\ref{sec:methodsorderparameters}). The resulting order parameter vanishes above the N{\'e}el temperature and saturates at its ideal value at low temperature (see {\suppSecDirector}). Because it probes third-rank correlations of the individual spins, it is suppressed much more strongly by thermal fluctuations than the projection coefficient in Fig.~\ref{fig:groundstate}c. The spatial toroidal {\TfnOneMultipole} projection additionally resolves the assignment of these directions to crystallographic sites, providing a more specific characterization of the magnetic texture. The ten-spin-loop directors introduced for GGG~\cite{Paddison2015} describe the axes of alternating spin sums around the shortest non-triangular loops of the hyperkagome lattice (see Section~\ref{sec:methodsorderparameters}), i.e.~these characterize short-range order on ten-rings within the hyperkagome lattice. In almandine, these directors become fully correlated in the ordered phase, which can be traced as a function of separation along one-dimensional chains of loop centers (see {\suppSecDirector}). The loop directors thus provide a common description of collective spin organization in GGG and almandine. In GGG, their long-range correlations characterize hidden order within a spin-liquid state without conventional long-range spin order~\cite{Paddison2015}. In almandine, the ideal director pattern follows directly from the spatial {\TfnOneMultipole} texture through the alternating loop sums and therefore accompanies conventional non-collinear magnetic order. At finite temperature, the director correlations additionally characterize spatial fluctuations that are not determined by the coherent toroidal amplitude alone. The staggered spin tensor, the spatial toroidal projection, and the loop directors thus characterize complementary aspects of the same non-collinear magnetically ordered state.

\begin{figure*}
\centering
\includegraphics[width=\linewidth]{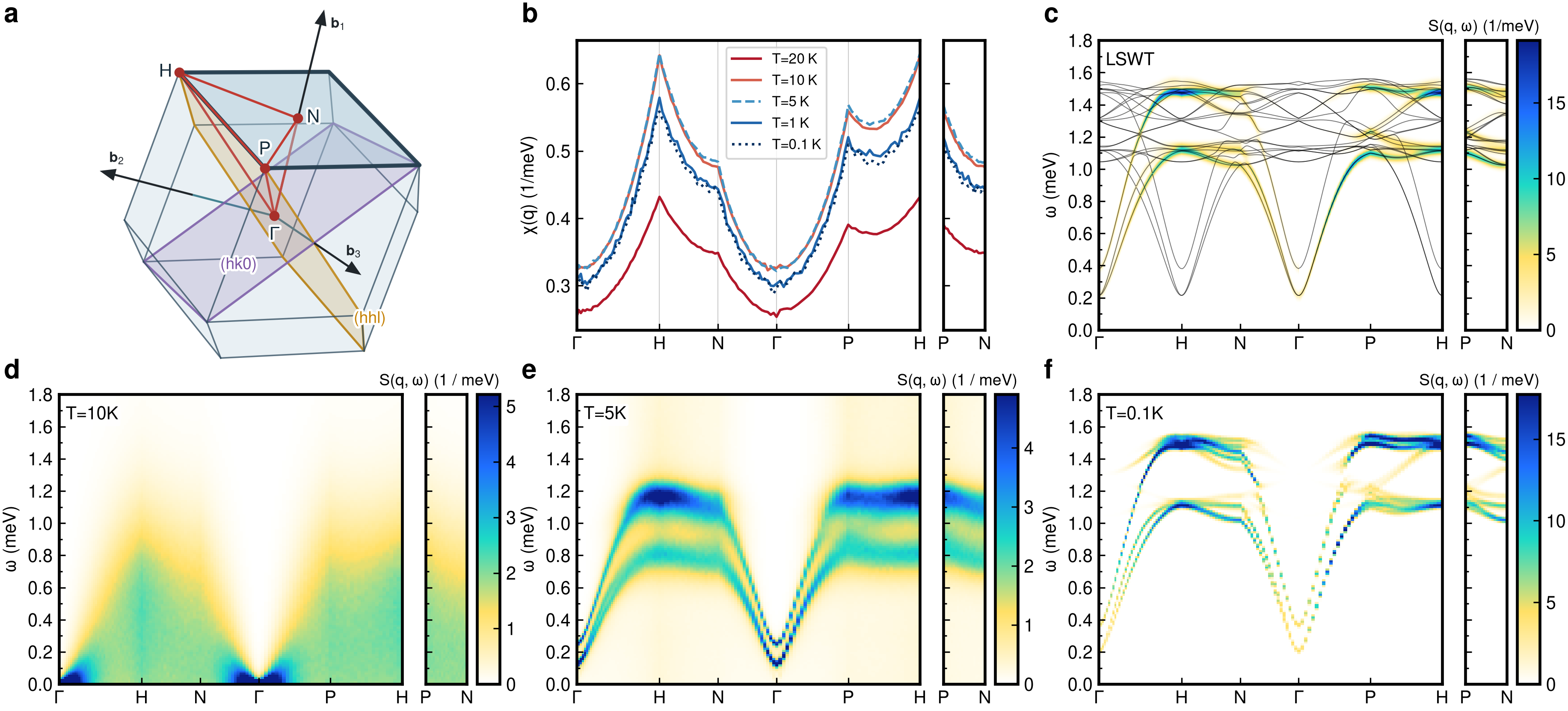}
\caption{\textbf{Static susceptibility and dynamical structure factor of almandine on a high-symmetry path.} \textbf{a} Brillouin zone of almandine with $(hhl)$ and $(hk0)$ planes and a path containing the high-symmetry points~\cite{Setyawan2010}. \textbf{b} Static susceptibility of almandine from classical Monte Carlo as a function of temperature. \textbf{c} Dynamical structure factor from linear spin-wave theory (LSWT). \textbf{d-f} Dynamical structure factor from classical Monte Carlo (cMC) + semiclassical dynamics at fixed temperature. The cMC calculations were performed including dipole-dipole interactions in a $40 \times 40 \times 40$ supercell.}
\label{fig:ssf_dsf}
\end{figure*}

We also calculated single crystal neutron diffraction amplitudes based on the magnetic structure in Table~\ref{tab:magneticstructure} using the software package \texttt{Sunny.jl}~\cite{Dahlbom2025}. For more details, see Section~\ref{sec:methodsdiffraction}. Since the experimental neutron diffraction amplitudes were measured on a natural crystal of almandine with an estimated Fe-site occupation of only $c = 0.795$~\cite{Prandl1971}, we also scaled our calculated amplitudes accordingly. In a single crystal experiment, the scattering amplitudes are sensitive to the orientation of the spins. In the magnetic structure described above, the spins are constrained to a plane perpendicular to the local $[110]$ axis on each Fe site. Within this plane, our spin axes are rotated by $36.5^\circ$ (Fe1) and $35.0^\circ$ (Fe2) w.r.t.~the local $[100]$ axes given in Table~7 in Ref.~\cite{Prandl1971}. Nevertheless, our results agree well with the earlier calculation in Ref.~\cite{Prandl1971}. The agreement of both calculations with experiment is not perfect, probably for various experimental reasons described in Ref.~\cite{Prandl1971}. Fig.~\ref{fig:groundstate}d compares the experimental neutron diffraction amplitudes for natural almandine with an estimated iron occupancy of $c= 0.795$~\cite{Prandl1971} to our theoretical calculations using full ($c=1$) and reduced iron occupancy comparable to natural almandine. For the magnetic structure in the cubic cell used for the indexing and the numerical values for the neutron diffraction amplitudes, see {\suppSecStructure}.

Fig.~\ref{fig:groundstate}e and \ref{fig:groundstate}f show the magnetic susceptibility and specific heat of almandine obtained from cMC for several system sizes. The susceptibility exhibits a maximum around 6.5\,K, while the specific heat also displays a pronounced peak at approximately 6.5\,K. Both quantities show only a mild system size dependence (see inset of Fig.~\ref{fig:groundstate}f). With the spin-length rescaling~\cite{Mendili2026} discussed in Section~\ref{sec:methodscmc}, this corresponds to $T_{\rm N}\approx 6.5\,\mathrm{K}\, \cdot (S+1)/S \approx 9.8$\,K with $S=2$. Heat capacity anomalies associated with antiferromagnetic ordering were reported at 8.7\,K~\cite{Anovitz1993} and 9.2\,K~\cite{Dachs2012}, close to the rescaled value. The lower value of 7.5\,K from magnetic susceptibility~\cite{Prandl1971} was obtained for a natural crystal with an estimated Fe-site occupation of only 79.5\,\%. The simulations also reproduce the shape of the lambda-type anomaly in the specific heat measured in various natural and synthetic samples of almandine~\cite{Dachs2012}. Note that the classical specific heat approaches $k_{\rm B}$ per spin for $T\to0$, so the comparison is restricted to the vicinity of the transition. M\"ossbauer spectroscopy indicates that the N{\'e}el temperature of almandine is smeared out between 9.0 and 10.4\,K~\cite{Murad1987}. Our susceptibility is very similar in shape to the $M/H$ magnetization (divided by field) curve for an applied field of $H=2$\,kOe reported by de Oliveira {\it et al.}~\cite{deOliveira1989} for pure almandine with a N{\'e}el temperature of $10.0(3)$\,K. 

The magnetic point group $\bar{3}'m$ of this state contains spatial inversion only in combination with time reversal, so a linear magnetoelectric (ME) effect is symmetry-allowed. The toroidal dipole $T_z$ along the trigonal axis belongs to the same $A_{2u}$ sector as the {\TfnOneMultipole} octupole and is the order parameter of the antisymmetric ME tensor derived below. Its point-dipole lattice sum vanishes for the classical moment texture (see Section~\ref{sec:methodsmultipole}), so symmetry fixes the form of the response but not its magnitude. The linear ME effect is described by the free-energy term~\cite{Fiebig2005} 
\begin{equation}
F=-\alpha_{ij}E_iH_j,\quad P_i=\alpha_{ij}H_j,\quad \mu_0M_j=\alpha_{ij}E_i .
\label{eq:F}
\end{equation}
$\alpha$ couples a polar, time-even vector ($\bm E$) to an axial, time-odd one
($\bm H$).

The symmetries of the magnetic state we determined earlier constrain the form of the linear magnetoelectric tensor to
\begin{equation}
\alpha=\tau\begin{pmatrix}0&1&0\\-1&0&0\\0&0&0\end{pmatrix},
\qquad
\alpha_{ij}=\tau\,\epsilon_{ijk}\,n_k ,
\label{eq:alpha}
\end{equation}
with a single independent coefficient $\tau$. In cubic axes, for
$\nh=[111]/\sqrt3$,
\begin{equation}
\alpha=\frac{\tau}{\sqrt3}\begin{pmatrix}0&1&-1\\-1&0&1\\1&-1&0\end{pmatrix}.
\label{eq:alphacubic}
\end{equation}

\begin{figure*}
\includegraphics[width=1.0\linewidth]{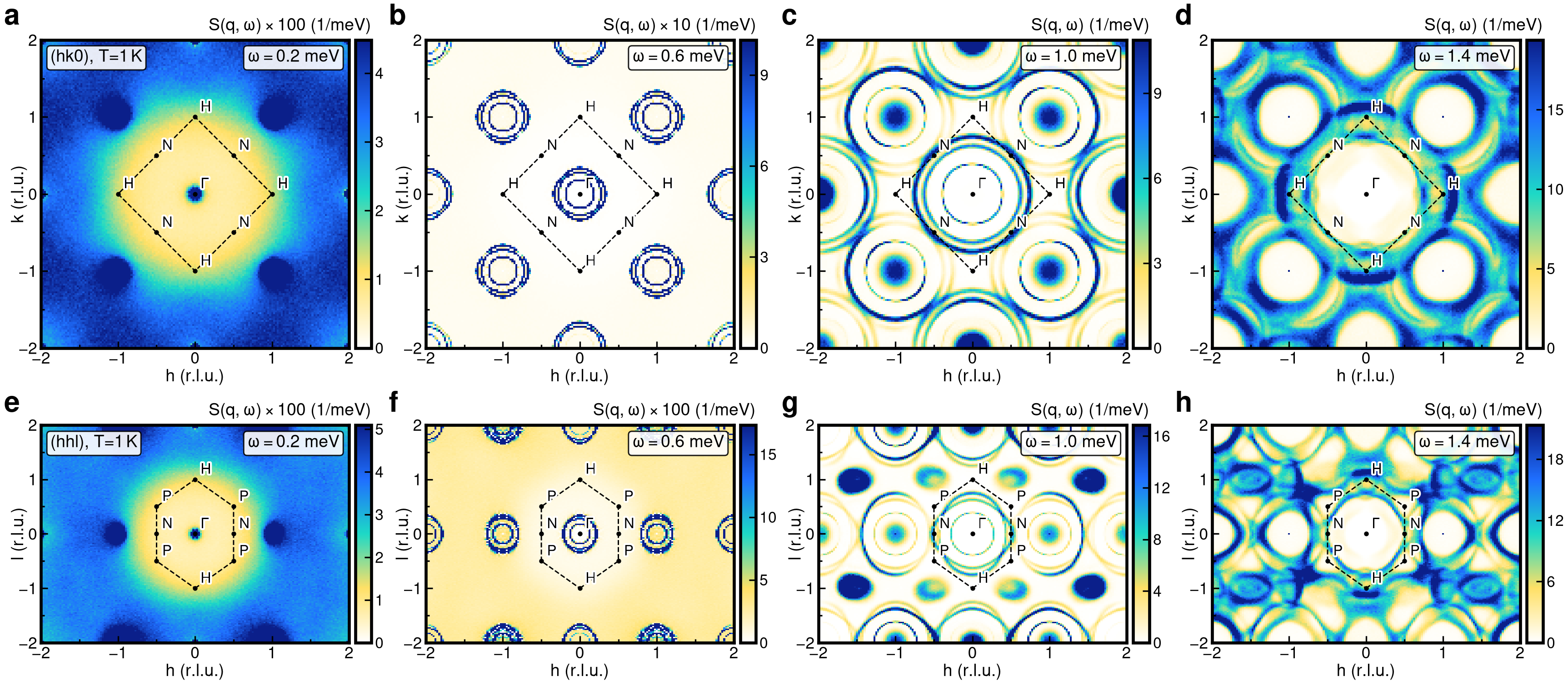}
\caption{\textbf{Dynamical structure factor of almandine at $\mathbf{T=1}\,\mathbf{K}$ and at constant excitation energy $\boldsymbol{\omega}$ in two-dimensional planes.} \textbf{a-d} Dynamical structure factor from classical Monte Carlo + semiclassical dynamics in the $(hk0)$ plane and \textbf{e-h} in the $(hhl)$ plane. All calculations were performed including dipole-dipole interactions in a $40 \times 40 \times 40$ supercell.}
\label{fig:dsfcut1K}
\end{figure*}

With $\bm\tau\equiv\tau\nh$, Eq.~\eqref{eq:F} becomes
\begin{equation}
\bm P=\bm H\times\bm\tau,\quad
\mu_0\bm M=\bm\tau\times\bm E,\quad
F=-\bm\tau\cdot(\bm E\times\bm H) \,.
\label{eq:vector}
\end{equation}
$\bm P$ is perpendicular to both $\bm H$ and $\nh$. There is no response to
$\bm H\parallel\nh$ and no polarization along the applied field. Defining
$\alpha$ through $\mu_0M_i=\alpha_{ij}E_j$ instead transposes it and reverses
the sign of $\tau$.

Since $F=-\bm\tau\cdot(\bm E\times\bm H)$, cooling through $T_{\rm N}$ in crossed electric and magnetic fields selects one of the eight domains (four orientations of the trigonal axis, each with $\pm\bm\tau$). Without such magnetoelectric annealing $\alpha$ averages to zero in a cubic crystal. The form of $\alpha$ also tests the spin orientation: the alternative $R\bar{3}'c'$ structure (magnetic point group $\bar{3}'m'$, as in Cr$_2$O$_3$~\cite{Fiebig2005}) forbids $T_z$ and would yield a diagonal tensor. The antisymmetric tensor we find above is consistent with reports of a sign-changing linear magnetoelectric effect in the similar garnet {\mnge}~\cite{Min2022}. A more recent report, however, finds a different point group and more independent components of the linear magnetoelectric tensor~\cite{Tan2024}. Here, we cannot comment on the situation in {\mnge}, but it seems that more aluminosilicate and -germanate materials in the garnet structure could show a linear magnetoelectric effect.

Next we use linear spin-wave theory (LSWT, see Section~\ref{sec:methodslswt})~\cite{Dahlbom2025} and cMC + semiclassical dynamics~\cite{Remund2022} to calculate the static susceptibility and dynamical structure factor of almandine to predict the outcome of future neutron experiments. The cubic Brillouin zone with high-symmetry points and planes is shown in Fig.~\ref{fig:ssf_dsf}a. The static susceptibility was calculated by extracting the static correlator from the semiclassical dynamics simulations. Fig.~\ref{fig:ssf_dsf}b shows the static susceptibility of almandine on the high-symmetry path~\cite{Setyawan2010} as a function of temperature. Clearly, the dominant peak is at the H point with subdominant humps of similar magnitude at N and P.

Linear spin wave theory (see Fig.~\ref{fig:ssf_dsf}c) based on the ordered antiferromagnetic ground state shows a gap of about 0.2\,meV in the dynamic structure factor, which is induced by dipole-dipole interactions. The dynamical structure factor without dipole-dipole terms is gapless; see {\suppSecDSF}. The magnon spectrum shows its minimum excitation energy at the $\Gamma$ point and maxima at H, N and P. Between the latter three points the dispersion is almost flat, indicating the magnetic frustration we expected from the Hamiltonian we determined above.

Classical Monte Carlo + semiclassical dynamics (for details see Section \ref{sec:methodscmc}) allows us to calculate the dynamical structure factor at finite temperature. Thus, Fig.~\ref{fig:ssf_dsf}d-f show the evolution from $T=10$\,K (above the cMC ordering temperature) to $T=5$\,K (just below the ordering temperature) and down to $T=0.1$\,K  (deep in the ordered phase). As expected, the spectrum is incoherent above the ordering temperature, with the bulk of the spectral weight concentrated near $\Gamma$. Nevertheless, the dynamical structure factor already shows some incoherent high-energy background weight at the H, N and P points. At $T=5$\,K we see a coherent excitation spectrum with substantial thermal broadening and a small gap of about $0.1$\,meV at the $\Gamma$ point. At $T=0.1$\,K the semiclassical dynamics results are close to LSWT, with a similar gap of about $0.2$\,meV. As in the LSWT case, the spectrum of cMC + semiclassical dynamics is gapless without dipole-dipole interactions (see {\suppSecDSF}).

In addition to the dynamical structure factor on the high-symmetry path of Fig.~\ref{fig:ssf_dsf}, we have also prepared constant-energy cuts in the $(hk0)$ and $(hhl)$ planes at constant temperature $T=1$\,K. The results in Fig.~\ref{fig:dsfcut1K} show a rich structure. As expected from Fig.~\ref{fig:ssf_dsf}, low-energy cuts show structure predominantly around the $\Gamma$ point, while higher excitation energies lead to greater complexity and a shift of the maxima in the spectral function to the H, N and P points. Thus, our results for the dynamical structure factor provide an experimentally testable prediction for the magnetic Hamiltonian and ground state we determined above.

\begin{table}[t]
\caption{\textbf{Hamiltonian parameters of gadolinium gallium garnet {\gagaga} (GGG) and almandine {\fe}.}  $d/a$ is the bond length in units of the cubic lattice constant, $z$ is the number of neighbors on this bond, and $J_2$, $J_3$ and $J_{4b}$ connect the two hyperkagome lattices. For GGG, $J_1$, $D$, and the best-fit values of $J_2$ and $J_3$ are taken from Ref.~\cite{Yavorskii2006}, whose error bars correspond to $-0.112\leq J_2/J_1\leq -0.037$ and $-0.028\leq J_3/J_1\leq 0.112$; couplings beyond $J_3$ are not part of the GGG Hamiltonian. For almandine, $D=\mu_0(g\mu_{\rm B})^2/(4\pi k_{\rm B}r_{\rm nn}^3)$ is evaluated for the nearest-neighbor distance $r_{\rm nn}=3.521$\,\AA\ with $g=2$; the same expression yields the GGG value of $D$. The values of $\theta_{\rm CW}$ are experimental results for GGG~\cite{SackvilleHamilton2014} and almandine~\cite{deOliveira1989}; the latter is reproduced by our couplings.}
\label{tab:gggcomparison}
\begin{ruledtabular}
\begin{tabular}{lccc}
& $d/a$ ($z$) & GGG & {\fe} \\
\hline
$S$ & & $7/2$ & $2$ \\
$\theta_{\rm CW}$ (K) & & $-2.6(1)$ & $-34$ \\
$J_1$ (K) & $0.306$ ($4$) & $0.107$ & $2.30$ \\
$D$ (K) & & $0.0457$ & $0.0571$ \\
\hline
$D/J_1$ & & $0.427$ & $0.0248$ \\
$J_2/J_1$ & $0.468$ ($8$) & $-0.047$ & $0.118$ \\
$J_3/J_1$ & $0.500$ ($2$) & $0.093$ & $0.122$ \\
$J_{4b}/J_1$ & $0.559$ ($4$) & --- & $0.377$ \\
\end{tabular}
\end{ruledtabular}
\end{table}

It is also instructive to compare our Hamiltonian to the prototypical hyperkagome material gadolinium gallium garnet (GGG), {\gagaga}. Both can be described by short-range Heisenberg exchange and a dipole-dipole interaction~\cite{Yavorskii2006}
\begin{equation}
\begin{aligned}
&H=\sum_{i<j}J_{ij}\,{\bf S}_i\cdot{\bf S}_j\\
&\quad+D\,r_{\rm nn}^3\sum_{i<j}\left[{\bf S}_i\cdot{\bf S}_j-3({\bf S}_i\cdot\hat{\bf r}_{ij})({\bf S}_j\cdot\hat{\bf r}_{ij})\right]/r_{ij}^3\,\,,
\end{aligned}
\end{equation}
with every bond counted once. 

With this common Hamiltonian convention, the dimensionless coupling ratios for GGG and almandine are directly comparable (Table~\ref{tab:gggcomparison}). The much larger nearest-neighbor exchange $J_1$ in almandine makes dipolar interactions relatively weaker and is reflected in the larger magnitude of its Curie-Weiss temperature $|\theta_{\rm CW}|$. Although $|J_2|/J_1$ and $|J_3|/J_1$ are of similar magnitude in the two compounds, almandine is distinguished by its substantial antiferromagnetic inter-hyperkagome coupling $J_{4b}$. This interaction favors antiparallel alignment of the spins it connects while allowing every nearest-neighbor triangle to retain a coplanar $120^\circ$ arrangement, thereby restricting the ground-state manifold without increasing the nearest-neighbor exchange energy. In GGG, dipolar interactions also couple the two hyperkagome lattices. These dipole-dipole interactions are themselves frustrated and generate local magnetic anisotropy. Their interplay with nearest- and further-neighbor exchange produces a delicate competition between magnetic states~\cite{Yavorskii2006}.

In almandine this degeneracy is clearly lifted by $J_{4b}$: as a thought experiment, we have calculated the magnetic specific heat, magnetic susceptibility, and dynamical structure factor of a reduced $J_1$-$J_{4b}$ model (see {\suppSecMinimalModel}) with and without dipole-dipole interactions. Reducing $J_{4b}$ decreases the ordering temperature of this model until no order can be detected at $J_{4b} =0$. For $J_{4b} > 0$ the ground state order remains the same as in the full model we determined for almandine, as expected from our previous analysis. At $J_{4b}=0$ the degeneracy of the ground state is clearly visible in the incoherent dynamical structure factor, which shows a large amount of low-energy weight near the H point, attributable to a manifold of nearly degenerate low-energy states. After adding a dipole-dipole interaction, these low-energy states are gapped out by the dipole-dipole interaction at sufficiently low temperature, which freezes a disordered state upon cooling.

\section{Discussion}
Our results show that the natural mineral almandine {\fe} realizes a hyperkagome antiferromagnet. We have determined its Heisenberg Hamiltonian using density functional theory energy mapping, which yields significant longer range couplings, namely the $J_{4b}$ inter-hyperkagome coupling, which determines the ground state order. Classical Monte Carlo simulations of this Hamiltonian give magnetic specific heat and susceptibility in agreement with experiments, including the N{\'e}el temperature once the classical spin length is rescaled to $\sqrt{S(S+1)}$~\cite{Mendili2026}. Furthermore, the diffraction pattern based on the ground state from our cMC simulations is consistent with early neutron diffraction experiments. Our simulations for the static and dynamical structure factor of almandine provide detailed testable predictions for future neutron scattering experiments, where dipole-dipole interactions open a small gap.

Our results differ qualitatively from an earlier attempt by Zherebetskyy {\it et al.}~\cite{Zherebetskyy2012} to determine the exchange couplings of almandine from local spin density calculations on finite clusters. They found a ferromagnetic nearest-neighbor coupling, which they attributed to the Fe-O-Fe bond angle of about $101^\circ$, and a weak antiferromagnetic coupling between the two hyperkagome lattices mediated by Al octahedra and Si tetrahedra. The latter is consistent with our antiferromagnetic inter-hyperkagome couplings $J_2$ and $J_{4b}$, both of which connect Fe sites sharing an Al octahedron. Our $J_1$, however, is antiferromagnetic for all values of $U$ we considered and grows as $U$ decreases (Fig.~\ref{fig:couplings}a), so its sign does not depend on the choice of $U$. The couplings of Ref.~\cite{Zherebetskyy2012} imply an unfrustrated collinear state, in which each hyperkagome lattice orders ferromagnetically and the two lattices align antiparallel. This state is difficult to reconcile with experiment. First, the large negative Curie-Weiss temperature $\theta_{\rm CW}=-34$\,K indicates dominant antiferromagnetic exchange~\cite{deOliveira1989}. Second, the magnetic structure factor of a ferromagnetically ordered hyperkagome lattice vanishes identically at the $(310)$, $(510)$, $(301)$ and $(501)$ reflections (see {\suppSecNeutronFM}). This holds for any spin direction and any relative orientation of the two lattices. Prandl, however, observed purely magnetic intensity at all four reflections, with $(310)$ among the strongest~\cite{Prandl1971}. Our non-collinear ground state yields sizable amplitudes at these reflections (see Fig.~\ref{fig:groundstate}d and {\suppSecStructure}).

Furthermore, we could show that the magnetic ground state of almandine is a magnetic toroidal octupolar order of {\TfnOneMultipole} type. The linear magnetoelectric tensor of almandine is, therefore, antisymmetric with a single independent coefficient, consistent with experimental reports of a sign-changing linear magnetoelectric effect in the similar garnet {\mnge}.

We also compared our Hamiltonian for almandine with that of the prototypical gadolinium gallium garnet {\gagaga}. Our results identify the substantial antiferromagnetic inter-hyperkagome coupling $J_{4b}$ as the main interaction promoting conventional magnetic order in almandine. Its bonds can be satisfied by antiparallel spin alignment while preserving the $120^\circ$ arrangement on every nearest-neighbor triangle, thereby restricting the frustrated ground-state manifold. Dipolar interactions are relatively weak compared with almandine's much larger exchange scale $J_1$. In GGG, the relatively strong dipolar interactions also couple the two hyperkagome lattices, but are themselves frustrated and generate local anisotropy. Their competition with exchange explains the different ordering tendencies of the two compounds.

Our finding of strongly frustrated hyperkagome antiferromagnetism in the natural mineral almandine encourages further research into the family of aluminosilicates and their germanate analogs in the garnet structure. It seems likely that this family of materials may contain many further realizations of hyperkagome antiferromagnets, which have not yet been discovered.

\section{Methods}
\subsection{Density functional theory-based energy mapping}
\label{sec:methodsdft}
We perform density functional theory based energy mapping~\cite{Guterding2016,Iqbal2018,Jaubert2025} using the full potential local orbital (FPLO)~\cite{Koepernik1999} basis, generalized gradient approximation (GGA)~\cite{Perdew1996} exchange correlation functional and DFT+U corrections~\cite{Liechtenstein1995} to deal with the strong electronic correlations on Fe $3d$ orbitals. We fix the Hund's rule coupling to the values proposed in Ref.~\onlinecite{Mizokawa1996}, i.e.~$J_{\rm H}=0.84$\,eV for Fe. We fix $U$ by evaluating the Curie-Weiss temperature in the high-temperature expansion for quantum spins, $\theta_{\rm CW}=-\frac{S(S+1)}{3 k_B}\sum_n z_nJ_n$~\cite{Johnston2015}, where $z_n$ is the number of neighbors coupled by $J_n$.

We double the primitive unit cell of almandine and calculate 40 distinct spin configurations of the 24 inequivalent Fe sites in order to work out seventeen inequivalent exchange interactions up to about three times the nearest neighbor distance.

\subsection{Classical Monte Carlo and semiclassical dynamics}
\label{sec:methodscmc}
We performed cMC simulations for the Hamiltonian in Eq.~\ref{eq:H} with and without a parameter-free dipole-dipole interaction using the Metropolis-Hastings algorithm~\cite{Metropolis1953, Hastings1970} with local updates~\cite{Venkatasubramanian2026,Jeschke2026}. The random spins for local updates are chosen according to a cone-based adaptive rule~\cite{AlzateCardona2019}, which keeps the acceptance rate of the Metropolis algorithm close to 50\%. After each Metropolis lattice sweep, we perform one energy-conserving overrelaxation sweep~\cite{Kadena1994} for the entire lattice. 

We use a supercell of size $L \times L \times L$. Since the doubled unit cell of almandine has 24 basis sites, this results in $24 L^3$ Fe sites in total. The dipole-dipole interaction was approximated by discarding terms beyond the twelfth neighbor for each Fe site for performance reasons. Thermodynamic quantities were estimated using 10000 warmup sweeps and 10000 subsequent sweeps for the measurement. Samples were averaged over 512 parallel runs.

In the cMC simulations, the spins are classical vectors of length $S=2$. For this normalization, the classical high-temperature expansion yields $\theta_{\rm CW}^{\rm cl}=\frac{S}{S+1}\,\theta_{\rm CW}\approx-23$\,K instead of $-34$\,K, i.e.\ all classical temperature scales are reduced by the factor $S/(S+1)=2/3$ relative to the quantum spin model. For large $S$, the partition function of the quantum model approaches that of a classical model with spin length $\sqrt{S(S+1)}$, and classical simulations with spin length $S$ correspondingly underestimate transition temperatures by $S/(S+1)$~\cite{Mendili2026}. We quote all cMC temperatures for spin length $S$. For comparison with experiment, they should be multiplied by $(S+1)/S=3/2$. The semiclassical dynamics and LSWT also use spin length $S$, which reproduces the spin-wave energies to leading order in $1/S$. Therefore, we do not adjust the energy scale of the structure factors.

In the semiclassical dynamics~\cite{Pohle2021,Remund2022} calculations we take a configuration from the cMC run and perform a classical time-evolution with the Hamiltonian in Eq.~\ref{eq:H}. For the time-evolution we rewrite the Hamiltonian using normalized spins $|\mathbf s_i|=1$ with $\mathbf S_i=S\mathbf s_i$:
\begin{equation}
H
= \frac{1}{2}\sum_{i,j}J_{ij}\,
\mathbf S_i\cdot\mathbf S_j
= \frac{S^2}{2}\sum_{i,j}J_{ij}\,
\mathbf s_i\cdot\mathbf s_j \,.
\end{equation}

Here, $J_{ij}=J_{ji}$ and the sum counts both orientations of each bond. We use natural units with $\hbar=1$ for the semiclassical dynamics. The spins are dimensionless, and our time variable $t=t_{\mathrm{phys}}/\hbar$ is measured in inverse-energy units. Throughout the dynamical calculations and figures, $\omega$ denotes the excitation energy $\hbar\Omega$, where $\Omega$ is the physical angular frequency. Thus, when energies are expressed in meV, time is expressed in $\mathrm{meV}^{-1}$. Exchange couplings quoted in kelvin represent $J_{ij}/k_{\mathrm B}$. The semiclassical equations of motion are
\begin{equation}
\frac{d\mathbf S_i}{dt}
= -\mathbf S_i\times
\frac{\partial H}{\partial\mathbf S_i}
= -\mathbf S_i\times\sum_j J_{ij}\mathbf S_j \,.
\end{equation}
Equivalently, the normalized spins evolve according to
\begin{equation}
\frac{d\mathbf s_i}{dt}
= -S\,\mathbf s_i\times\sum_j J_{ij}\mathbf s_j \,.
\end{equation}
We integrate the resulting system of ordinary differential equations with the Dormand-Prince algorithm~\cite{Dormand1980}. Dormand-Prince is an adaptive method, i.e.~we set a small acceptable maximum error and the algorithm will decrease the width of time steps where necessary and use larger time steps where possible. In cMC runs with dipole-dipole interactions, these are also included in the semiclassical dynamics.

The dynamical structure factor $S_\text{cl}({\bf q}, \omega; T)$ is estimated from the sampled trajectory. For each trajectory $r=1,\ldots,R$, we store the spins at
$t_n=n\Delta t$, with $n=0,\ldots,N_t-1$, and define
\begin{equation}
\mathbf F_r(\mathbf q,t_n)
= \frac{1}{\sqrt N}\sum_{i=1}^{N}
\mathbf s_{i,r}(t_n)e^{i\mathbf q\cdot\mathbf r_i} \,.
\end{equation}
The time correlator is estimated as
\begin{equation}
\begin{aligned}
&C(\mathbf q,m\Delta t) \\
&= \frac{1}{R}\sum_{r=1}^{R}
\frac{1}{N_t-m}\sum_{n=0}^{N_t-m-1}
\mathbf F_r(\mathbf q,t_{n+m})
\cdot\mathbf F_r(\mathbf q,t_n)^* \,, \\
&0\leq m\leq N_t-1 \,.
\end{aligned}
\end{equation}
Negative lags satisfy
\begin{equation}
C(\mathbf q,-m\Delta t)
= C(\mathbf q,m\Delta t)^* \,.
\end{equation}
The static structure factor is obtained as
\begin{equation}
S(\mathbf q)
= \operatorname{Re}\, C(\mathbf q,0)
= \frac{1}{N}\sum_{i,j}
\left\langle \mathbf s_i \cdot \mathbf s_j \right\rangle
e^{i\mathbf q\cdot(\mathbf r_j-\mathbf r_i)} \,,
\label{eq:static-structure-factor}
\end{equation}
where $N$ is the number of spins, $\mathbf s_i$ are unit-length classical spin vectors, and $\mathbf r_i$ are their positions. The average is taken over time origins within each trajectory and over Monte Carlo sampled trajectories.

In Fig.~\ref{fig:ssf_dsf}b we show the classical wave-vector-dependent static susceptibility
\begin{equation}
\chi(\mathbf q) = \frac{S(\mathbf q)}{k_{\rm B}T} \,,
\label{eq:static-susceptibility}
\end{equation}
which follows from the classical fluctuation-dissipation theorem and is given in units of 1/meV. For unit-length spins it corresponds to the trace $\sum_\alpha\chi^{\alpha\alpha}(\mathbf q)$ per spin, without the prefactor $(g\mu_{\rm B}S)^2$. Dividing by $k_{\rm B}T$ removes the linear temperature dependence of the classical $S(\mathbf q)$ in the ordered phase, so curves at different temperatures can be compared directly.

The classical dynamical structure factor is obtained via the discrete time Fourier transform
\begin{equation}
\begin{aligned}
S_{\mathrm{cl}}(\mathbf q,\omega_k) &= \Delta t\sum_{m=-(N_t-1)}^{N_t-1}
   w_m C(\mathbf q,m\Delta t)
   e^{i\omega_k m\Delta t} \,, \\
\omega_k &= \frac{2\pi k}{(2N_t-1)\Delta t} \,,
\end{aligned}
\end{equation}
where $k=0,\ldots,N_t-1$ denotes the nonnegative-energy
output branch. The window function is
\begin{equation}
w_m
= \frac{1}{2}\left[
1+\cos\left(\frac{\pi |m|}{N_t-1}\right)
\right] \,.
\end{equation}
However, the above algorithm only gives us classical occupation probabilities for excited states. To recover an approximation for the quantum-mechanical dynamical structure factor comparable, e.g.~to LSWT, we apply an energy-dependent Bose correction factor~\cite{Remund2022}:
\begin{equation}
S_\text{QM} ({\bf q}, \omega; T) \simeq \frac{ \omega}{k_B T \left( 1 - e^{- \omega / k_B T} \right)} S_\text{cl}({\bf q}, \omega; T)
\, .
\end{equation}
All dynamical structure factors show this corrected form.

Static and dynamical structure factors were estimated using 30000 warmup sweeps followed by at least 300 subsequent sweeps for the measurement, with a semiclassical time evolution performed every 10 lattice sweeps. Samples were then averaged over 48 parallel runs.

\subsection{Simulation of neutron diffraction}
\label{sec:methodsdiffraction}
We calculated single-crystal magnetic neutron-scattering amplitudes by direct summation over the ordered Fe moments, using \texttt{Sunny.jl}~\cite{Dahlbom2025} to construct the magnetic structure from the data in Table~\ref{tab:magneticstructure} and evaluate the tabulated Fe$^{2+}$ magnetic form factor. We assumed spin-only Fe$^{2+}$ moments with $S=2$ and $g=2$, corresponding to $4\,\mu_{\mathrm B}$ per Fe site. For each reflection, we evaluated the magnetic structure factor including the tabulated Fe$^{2+}$ magnetic form factor and the projection of the moments perpendicular to the scattering vector, which depends on the chosen $(hkl)$. The reported scattering amplitudes were obtained as $A_{\mathrm M}=p \cdot |\mathbf F_{\mathrm M,\perp}|$, where $p=2.695\,\mathrm{fm}/\mu_{\mathrm B}$ and $\mathbf F_{\mathrm M,\perp}$ is expressed in $\mu_{\mathrm B}$. For comparison with the natural-crystal measurements of Ref.~\cite{Prandl1971}, we multiplied the amplitudes by the Fe-site occupancy factor $c=0.795$. Magnetic Debye-Waller factors were set to unity.

\subsection{Linear spin wave theory}
\label{sec:methodslswt}
We calculated the dynamical spin structure factor using linear spin-wave theory as implemented in \texttt{Sunny.jl}~\cite{Dahlbom2025}. We supplemented the previously determined Heisenberg Hamiltonian with long-range dipole--dipole interactions, whose strength is fixed by the crystal geometry and the spin-only Fe$^{2+}$ moments with $S=2$ and $g=2$. These interactions therefore introduce no additional fitted parameter and were evaluated using Ewald summation. The spin-wave calculation was based on the ordered magnetic structure obtained from classical Monte Carlo simulations. We diagonalized the quadratic spin-wave Hamiltonian to obtain the magnon energies and their scattering weights. The zero-temperature neutron-scattering intensities include the tabulated Fe$^{2+}$ magnetic form factor and the projection perpendicular to the scattering vector. The discrete modes were broadened with a Gaussian of full width at half maximum $0.04\,\mathrm{meV}$.

\subsection{Multipole calculations}
\label{sec:methodsmultipole}
We use \texttt{MultiPie}~\cite{Kusonose2023} to determine the symmetry-allowed multipoles. The magnetic and magnetic-toroidal multipoles considered here are linear in the Fe magnetic moments and, therefore, odd under time reversal. The magnetic space group $R\bar{3}'c$ (BNS 167.105) has magnetic point group $\bar{3}'m$ (20.3.73)~\cite{Yatsushiro2021}. Invariance under its unprimed threefold rotations and mirrors requires the multipoles to remain unchanged under these spatial operations. Inversion and twofold rotations about axes perpendicular to the trigonal axis are combined with time reversal; the multipoles must therefore change sign under their spatial parts so that the combined operations leave them invariant. These transformation properties uniquely select the irreducible representation $A_{2u}$ of the ordinary trigonal point group $D_{3d}=\bar{3}m$. 

Up to rank five, the time-odd multipoles in this sector are the magnetic toroidal dipole $\mathbb{T}_{1}(A_{2u})$ with $h=z$, two magnetic toroidal octupoles, one magnetic hexadecapole and two rank-5 magnetic toroidal multipoles. The Fe moments of the magnetic cell, however, span only a four-dimensional symmetric subspace, because each of the two Fe orbits retains two free moment components perpendicular to its primed twofold axis. We therefore construct the basis in order of increasing rank and keep a multipole only if its symmetrized pattern $P\bm b_j$ (see below) is nonzero and linearly independent of the patterns already selected. The toroidal dipole drops out at this step: for the site images defined below, its pattern projects to zero, i.e.~no symmetric configuration on this lattice carries a unit-cell toroidal dipole $\frac12\sum_i\bm r_i\times\bm m_i$. $\mathbb{T}_{5}(A_{2u},1)$ is linearly dependent on the three lower-rank patterns. The remaining four polynomials $h_j$ span the symmetric subspace:
\begin{center}
\renewcommand{\arraystretch}{1.7}
\begin{tabular}{@{}ccll@{}}
\hline
$j$ & Rank $\ell$ & Multipole & Polynomial $h_j(x,y,z)$\\
\hline
1 & 3 & {\TfnOneMultipole} & $z^3-\frac32z(x^2+y^2)$\\
2 & 3 &  \(\mathbb{T}_{3}(A_{2u},2)\) & $\frac{\sqrt{10}}4y(3x^2-y^2)$\\
3 & 4 & \(\mathbb{M}_{4}(A_{2u})\) & $\frac{\sqrt{70}}4xz(x^2-3y^2)$\\
4 & 5 & \(\mathbb{T}_{5}(A_{2u},2)\) & $-\frac{\sqrt{70}}{16}y(3x^2-y^2)(x^2+y^2-8z^2)$\\
\hline
\end{tabular}
\end{center}
Here, $\mathbb{T}$ denotes magnetic toroidal multipoles and $\mathbb{M}$ magnetic multipoles and $n \in \{1,2 \}$ labels distinct harmonics. The coordinates $(x,y,z)$ are orthonormal Cartesian coordinates, with axes along cubic $[10\bar1]$, $[\bar12\bar1]$ and $[111]$, respectively. 

The multipole origin is the Al $16a$ site at cubic fractional coordinates $(0,0,0)$. For each Fe site, we choose the periodic image with cubic fractional coordinates in $[0,1)^3$. The vector $\bm r_i$ is the displacement from this Al site to the chosen Fe image, expressed in the Cartesian frame above. The multipole patterns and their individual weights depend on this choice of origin and site images. This position convention remains fixed during the global spin alignment described below.

The vector field $\bm b_j (\bm r)$ for each of the above multipole polynomials $h_j (\bm r) \equiv h_j (x,y,z)$ is defined as:
\begin{equation}
\bm{b}_j(\bm r)=
\begin{cases}
\bm\nabla h_j(\bm r),
& \text{magnetic }(\ell\geq 1),\\
\bm r\times\bm\nabla h_j(\bm r),
& \text{magnetic toroidal}.
\end{cases}
\label{eq:multipolevectorfield}
\end{equation}
This vector field represents a candidate spin pattern when evaluated at the magnetic sites in the unit cell.

The above multipoles are allowed by the magnetic \emph{point group}. The magnetic \emph{space group} also contains operations with fractional translations, such as glides. These operations map sites to other sites, possibly across a cell boundary. The polynomial vector field is not periodic, so its values at the chosen cell positions need not obey the required relations between the moments on those sites.

For multipole $j$, the vectors $\bm b_{j,i}=\bm b_j(\bm r_i)$ on the $N=24$ Fe sites form the candidate pattern $\bm b_j=(\bm b_{j,1},\ldots,\bm b_{j,N})$.
We average this pattern over the magnetic symmetry operations:
\begin{equation}
\bm p_j=P\bm b_j,
\qquad
P=\frac1{|G|}\sum_{g\in G}D(g).
\label{eq:multipoleaverage}
\end{equation}
Here $G$ contains the 24 operations acting on the conventional cell for BNS 167.105. If operation $g$ sends site $i$ to $\pi_g(i)$, its action is
\begin{equation}
\begin{aligned}
\bigl[D(g)\bm b_j\bigr]_{\pi_g(i)}
&=\epsilon_g\det(R_g)R_g\bm b_{j,i},\\
\epsilon_g
&=\begin{cases}
+1,&\text{without time reversal},\\
-1,&\text{with time reversal}.
\end{cases}
\end{aligned}
\label{eq:action}
\end{equation}
Here $R_g$ is the spatial rotation or reflection matrix in the Cartesian frame above. Translations enter through the site permutation. The determinant accounts for the axial character of magnetic moments.

The averaged pattern obeys the magnetic symmetry: $D(g)\bm p_j=\bm p_j$ for every $g\in G$. We normalize each nonzero pattern:
\begin{equation}
\bm e_{j,i}
=\frac{\bm p_{j,i}}
{\sqrt{\sum_{s=1}^{N}|\bm p_{j,s}|^2}}.
\end{equation}
For site-resolved patterns, we define
\begin{equation}
\langle\bm u,\bm v\rangle
=\sum_{i=1}^{N}\bm u_i\cdot\bm v_i,
\qquad
\|\bm u\|^2=\langle\bm u,\bm u\rangle.
\end{equation}
The normalized patterns $\bm e_j$ need not be mutually orthogonal. We therefore apply Gram--Schmidt in the order listed in the table:
\begin{equation}
\begin{aligned}
\bm u_j
&=\bm e_j-\sum_{k<j}
  \langle\bm q_k,\bm e_j\rangle\bm q_k,\\
\bm q_j&=\frac{\bm u_j}{\|\bm u_j\|}.
\end{aligned}
\end{equation}
The resulting patterns satisfy
$\langle\bm q_j,\bm q_k\rangle=\delta_{jk}$.

For each Monte Carlo configuration, we average the unit spins over the $N_c=12^3$ conventional cells:
\begin{equation}
\overline{\bm s}_i
=\frac1{N_c}\sum_{L=1}^{N_c}\bm s_{L,i}.
\end{equation}
The spins are expressed in the Cartesian frame defined above. We apply one global proper rotation $\mathcal R$, chosen to best align the averaged configuration with the reference texture in the conventional cell. The moments entering the projection are
\begin{equation}
\bm m_i=m_0\mathcal R\overline{\bm s}_i,
\qquad m_0=4\,\mu_{\rm B}.
\end{equation}

For these averaged and aligned magnetic moments, the signed projection coefficients and weights are
\begin{equation}
\begin{aligned}
c_j&=\sum_{i=1}^{N}\bm q_{j,i}\cdot\bm m_i,\\
w_j&=\frac{c_j^2}{\sum_{i=1}^{N}|\bm m_i|^2}.
\end{aligned}
\end{equation}
The four patterns span the symmetry-compatible configurations,
so
\begin{equation}
P\bm m=\sum_{j=1}^{4}c_j\bm q_j.
\end{equation}
The weights sum to one when the actual state obeys the chosen magnetic symmetry. Individual weights depend on the chosen site images and orthogonalization order.

Fig.~\ref{fig:groundstate}c shows the mean sublattice magnetization, the normalized coefficient of the {\TfnOneMultipoleAbbrev} pattern ($j=1$ in the table above), and its relative weight:
\begin{equation}
\begin{aligned}
|M_\alpha|
&=\frac1N\sum_{i=1}^{N}|\overline{\bm s}_i|,\\
c_{{\scriptscriptstyle\mathbb{T}}(3,1)}
&=\frac{c_1}{m_0\sqrt N},\\
w_{{\scriptscriptstyle\mathbb{T}}(3,1)}
&=w_1=\frac{c_1^2}{\sum_{i=1}^{N}|\bm m_i|^2}.
\end{aligned}
\end{equation}
The coefficient $c_{{\scriptscriptstyle\mathbb{T}}(3,1)}$ measures the signed amplitude of coherent octupolar order and satisfies $|c_{{\scriptscriptstyle\mathbb{T}}(3,1)}|\leq1$, with equality for the ideal {\TfnOneMultipole}
texture. The weight $w_{{\scriptscriptstyle\mathbb{T}}(3,1)}$ measures its relative contribution to the averaged configuration.

We set the displayed relative weight to zero when the coherent projection is small:
\begin{equation}
w_{{\scriptscriptstyle\mathbb{T}}(3,1)}^{\rm plot}=
\begin{cases}
w_{{\scriptscriptstyle\mathbb{T}}(3,1)},&\sqrt{\sum_{j=1}^{4}c_j^2}>2\,\mu_{\rm B},\\
0,&\text{otherwise}.
\end{cases}
\end{equation}

Fig.~\ref{fig:groundstate}b shows the polynomial of the dominant {\TfnOneMultipole} multipole centered on the Al site with Fe1 and Fe2 spins shown around it in different colors. 

\subsection{Alternative order parameters for the hyperkagome lattice}
\label{sec:methodsorderparameters}
Zhitomirsky~\cite{Zhitomirsky2008} characterizes orientational order using the symmetric traceless third-rank spin tensor. For classical unit spins $\mathbf{s}_i$, its local components are
\begin{equation}
t_i^{abc}
=s_i^a s_i^b s_i^c
-\frac{1}{5}\left(
s_i^a\delta_{bc}
+s_i^b\delta_{ac}
+s_i^c\delta_{ab}
\right) \,,
\end{equation}
where $a,b,c$ denote Cartesian spin components. The trace subtraction isolates the rank-three angular component of the spin-direction distribution. The instantaneous lattice-averaged tensor is
\begin{equation}
\mathcal{T}^{abc}=\frac{1}{N}\sum_i t_i^{abc} \,.
\end{equation}
The following scalar observable can be measured as a thermal average using Monte Carlo simulations,
\begin{equation}
\begin{aligned}
m_3^2
&=\left\langle
\sum_{abc}\mathcal{T}^{abc}\mathcal{T}^{abc}
\right\rangle\\
&=\frac{1}{N^2}\sum_{i,j}\left[
\left\langle(\mathbf{s}_i\cdot\mathbf{s}_j)^3\right\rangle
-\frac{3}{5}
\left\langle\mathbf{s}_i\cdot\mathbf{s}_j\right\rangle
\right] \,.
\end{aligned}
\end{equation}

In the ideal coplanar triatic state, spins throughout an ordered domain occupy a common set of three directions separated by $120^\circ$. The octupolar tensor measures the orientation of this set, independently of which site carries each direction. Equal populations of the three directions yield zero net magnetization while retaining a nonzero octupolar tensor. In the ideal limit,  the third-rank tensor gives $m_3^2=1/4$. Consequently, octupolar orientational order can exist without selecting a unique periodic magnetic structure.

In almandine, $J_{4b}$ lifts the ground-state degeneracy of the pure hyperkagome lattice, and the six low-temperature spin directions form two approximately opposite triads. The two sectors contribute opposite third-rank tensors, so their uniform sum vanishes in the ideal limit because $t^{abc}(-\mathbf{s})=-t^{abc}(\mathbf{s})$. We therefore use a staggered extension,
\begin{equation}
\mathcal{T}_{\mathrm{stag}}^{abc}
=\frac{1}{N}\sum_i\eta_i t_i^{abc} \,,
\end{equation}
where $\eta_i=\pm1$ are fixed site labels identifying the two triatic sectors of the reference texture. Its squared magnitude characterizes staggered triatic orientational order and is unchanged by permutations of spin directions among sites within each sector. The projection onto the spatial toroidal multipole $\mathbb{T}_3(A_{2u},1)$ additionally resolves the assignment of spin directions to crystallographic sites and generally changes under such permutations. The two observables thus characterize complementary aspects of the same non-collinear ordered state: the orientation of the triads and their spatial arrangement.

We also characterize collective spin correlations using the ten-spin-loop directors introduced by Paddison \emph{et al.}~\cite{Paddison2015}.
The loops are constructed from the nearest-neighbor $J_1$ connectivity of the hyperkagome networks and are the shortest closed paths beyond the elementary triangles. For each loop $\ell$, we define the alternating sum of its ten unit spins,
\begin{equation}
\mathbf{L}_{\ell}
=\frac{1}{10}\sum_{n=1}^{10}(-1)^n\mathbf{s}_{\ell n}
\,,
\qquad
\hat{\mathbf{L}}_{\ell}
=\frac{\mathbf{L}_{\ell}}{|\mathbf{L}_{\ell}|}
\,.
\end{equation}
Shifting the starting index by one site reverses $\mathbf{L}_{\ell}$, so its sign is a labeling convention. The physical director therefore represents an axis, with $\mathbf{L}_{\ell}$ and $-\mathbf{L}_{\ell}$ describing the same orientation. Its magnitude measures the alternating spin component within the loop, whereas the normalized director retains only its orientation.

Following Ref.~\onlinecite{Paddison2015}, we calculate the correlation function
\begin{equation}
g(r,T)
=2\left\langle
\left|\hat{\mathbf{L}}_{\ell}\cdot
\hat{\mathbf{L}}_{m}\right|
\right\rangle_{r,T}-1
\,,
\end{equation}
where the average includes cMC configurations at temperature $T$ and pairs of loops whose centers belong to the distance shell $r$. The pair correlation equals $+1$ for collinear axes and $-1$ for orthogonal axes. Its average vanishes for independent, isotropically distributed directors.

For the chains formed by loop centers connected at $r=5.7502\,\mathrm{\AA}$, we additionally evaluate
\begin{equation}
C(n,T)
=\left\langle
2\left|\hat{\mathbf{L}}_i\cdot
\hat{\mathbf{L}}_{i+n}\right|-1
\right\rangle
\,,
\end{equation}
where $n$ counts steps along a chain and the average includes all starting sites, equivalent chains, and cMC configurations. To quantify the geometrical background caused by overlapping loops, we repeat the director-correlation calculation for 2000 independent isotropic random-spin configurations with the same loop geometry. Further details and the resulting correlations are given in {\suppSecDirector}.

\section*{Data availability}
The data and scripts for all figures within this study are deposited at Zenodo~\cite{Zenodo}.

\section*{Code availability}
The DFT calculations in this study were performed using the FPLO package~\cite{Koepernik1999}, while the neutron diffraction and LSWT calculations were carried out using \texttt{Sunny.jl}~\cite{Dahlbom2025}. Multipole calculations were carried out with \texttt{MultiPie}~\cite{Kusonose2023}. The code used for the cMC and semiclassical dynamics calculations is available from the corresponding authors upon reasonable request.

\section*{Acknowledgments}
We acknowledge fruitful discussions with Bella Lake and Junya Otsuki. H.O.J. thanks for support through JSPS KAKENHI Grants No.~24H01668 and No.~25K08460. Part of the computation in this work has been done using the facilities of the Supercomputer Center, the Institute for Solid State Physics, the University of Tokyo. 
D.G.~acknowledges the support of the Investment Bank of the State of Brandenburg (ILB) under Grant No.~86001926, which was co-financed by the European Union through its regional development fund (ERDF).

\section*{Author contributions}
H.O.J.~conceived the project. D.G.~implemented and performed semiclassical dynamics calculations. D.G.~also carried out neutron diffraction and LSWT calculations. D.G.,~M.Z.~and H.O.J.~performed density functional theory calculations. All authors performed Monte Carlo calculations and analyzed the results. The manuscript was written with contributions from all authors.

\section*{Competing interests}
The authors declare no competing interests.

\bibliography{hyperkagome}

\clearpage
\includepdf[pages=1]{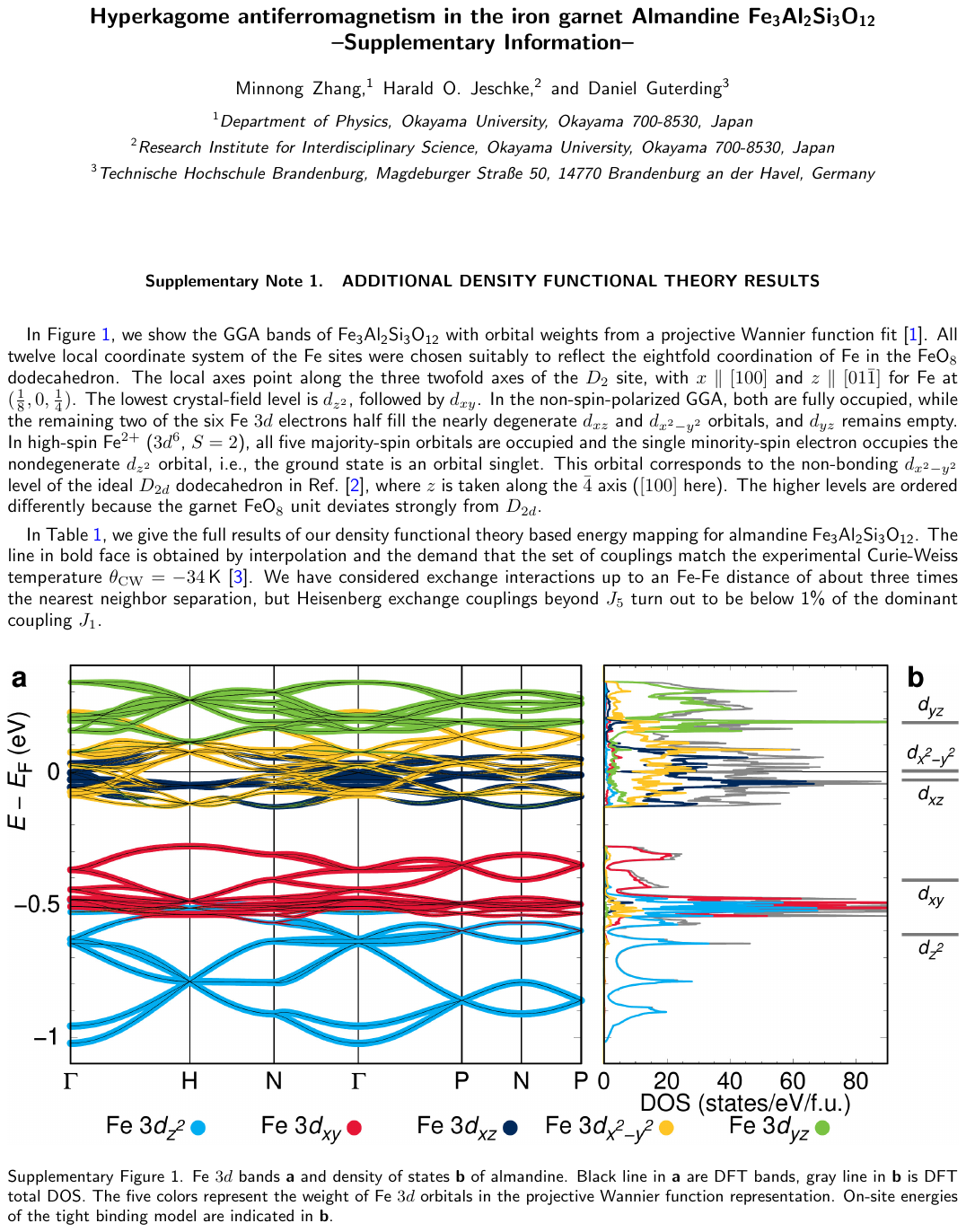}
\clearpage
\includepdf[pages=2]{supplement.pdf}
\clearpage
\includepdf[pages=3]{supplement.pdf}
\clearpage
\includepdf[pages=4]{supplement.pdf}
\clearpage
\includepdf[pages=5]{supplement.pdf}
\clearpage
\includepdf[pages=6]{supplement.pdf}
\clearpage
\includepdf[pages=7]{supplement.pdf}
\clearpage
\includepdf[pages=8]{supplement.pdf}
\clearpage
\includepdf[pages=9]{supplement.pdf}
\clearpage
\includepdf[pages=10]{supplement.pdf}
\clearpage
\includepdf[pages=11]{supplement.pdf}
\clearpage
\includepdf[pages=12]{supplement.pdf}
\end{document}